\documentclass[conference]{IEEEtran}
\usepackage{cite}
\usepackage{amsmath,amssymb,amsfonts}
\usepackage{algorithmic}
\usepackage{graphicx}
\usepackage{textcomp}
\usepackage{xcolor}

\def\BibTeX{{\rm B\kern-.05em{\sc i\kern-.025em b}\kern-.08em
		T\kern-.1667em\lower.7ex\hbox{E}\kern-.125emX}}

\newcommand\blankfootnote[1]{%
	\let\thefootnote\relax\footnotetext{#1}%
	\let\thefootnote\svthefootnote%
}

\begin{document}
	
	\title{Building Bridges: Establishing a Dialogue Between  Software Engineering Research and Computational Science}
	
	\author{\IEEEauthorblockN{Reed Milewicz}
		\IEEEauthorblockA{\textit{Department of Software Engineering and Research} \\
			\textit{Sandia National Laboratories}\\
			Albuquerque, NM \\
			rmilewi@sandia.gov}
	\and
	\IEEEauthorblockN{Miranda Mundt}
	\IEEEauthorblockA{\textit{Department of Software Engineering and Research} \\
	\textit{Sandia National Laboratories}\\
	Albuquerque, NM \\
	mmundt@sandia.gov}}
	
	\maketitle
	
	\begin{abstract}
		There has been growing interest within the computational science and engineering (CSE) community in engaging with software engineering research -- the systematic study of software systems and their development, operation, and maintenance -- to solve challenges in scientific software development. Historically, there has been little interaction between scientific computing and the field, which has held back progress. With the ranks of scientific software teams expanding to include software engineering researchers and practitioners, we can work to build bridges to software science and reap the rewards of evidence-based practice in software development.
	\end{abstract}
	
	\blankfootnote{Presented at the Workshop on the Science of Scientific-Software Development and Use, sponsored by U.S. Department of Energy, Office of Advanced Scientific Computing Research, Dec 13–15, 2021.
		
	Sandia National Laboratories is a multimission laboratory managed and operated by National Technology \& Engineering Solutions of Sandia, LLC, a wholly owned subsidiary of Honeywell International Inc., for the U.S. Department of Energy’s National Nuclear Security Administration under contract DENA0003525. SAND2021-14807 C.}
	
	
	\section{Challenge}\label{challenge}
	
	Software engineering research (or SE research) is the systematic study of software systems and their development, operation, and maintenance~\cite{stol2018abc}. A branch of computer science, software engineering research builds a bridge between the theory of computer programming and its practice as software engineering. It is likewise methodologically diverse, placing formal and empirical software analysis alongside human factors; it applies a sociotechnical lens to help model, understand, and predict the factors that lead to high-quality software systems. The field emerged in tandem with software engineering, with the first academic conferences and journals appearing in the mid-1970s. Since the beginning, there has been a symbiotic relationship between software engineering researchers and practitioners~\cite{osterweil2008determining};  as the science of software development, SE research provides a foundation for evidence-based practice in software engineering, and researchers work closely with industry to pioneer innovations in tools, techniques, and methodologies. Of importance to our work, there has been growing interest within the computational science and engineering (CSE) community in engaging with SE research to help address challenges in scientific software development~\cite{heroux2019softwarescience}.
	
	There has been little interaction between US DOE laboratories and the SE research community up until very recently, which has stymied the flow of innovation. Whereas the national labs have had a presence in HPC-related conferences and journals since the very beginning, we are virtually unknown in the realm of software engineering. Case in point, searching IEEE Xplore for publications by lab-affiliated authors at the ACM/IEEE International Conference on Software Engineering (ICSE) turns up just 7 articles since the conference was established in 1975 -- all of them in recent years. Likewise, IEEE Transactions on Software Engineering, one of the premier journals for SE research, has accepted exactly three articles from national lab authors in its history: one in 1984 and two in 1987. 	Current numbers of CSE participation in SE research are not readily available, but a 2002 review of the publication records of six prominent SE journals found that only 1\% of articles concerned computing in science and engineering~\cite{glass2002research}.
	
	%The opportunity cost of failing to engage substantively with software science is difficult to measure, but the benefits of doing so have not been lost on the tech industry. Silicon Valley firms actively court software engineering researchers and practitioners, inspiring the field to tackle their pressing challenges. They have established pipelines for talent from universities, have cultivated close ties with software scientists to remain on the cutting edge of innovation, and have consistently shown up to software engineering venues to advocate for their interests. If their success is any indication, computational science also stands to benefit from building bridges to software science.
	
	Meanwhile, because the CSE community has kept its distance, we are not yet conversant in software engineering literature and thus have yet to reap the rewards of evidence-based practice in software development. By evidence-based practice, we mean integrating current best evidence from research with practical experience and human values to improve decision-making related to software development and maintenance~\cite{kitchenham2004evidence}. The lack of familiarity with the science has had a direct impact on productivity and quality in CSE software development. As noted by the authors in a previous article~\cite{milewicz2020towards}, SE researchers have made great progress in understanding the factors that influence developer productivity with one recent meta-analysis identifying thirty five such factors~\cite{oliveira2018influence}. Each of those factors has been the subject of systematic, scientific investigation by SE researchers, and insights from those studies could inform CSE development practices. There may be situations where the applicability of studies is limited insofar as scientific software developers are underrepresented in the literature, but this is once again the result of a lack of engagement -- something which the CSE community can and should remedy.
	
	\section{Opportunity}
	
	%Describe how the identified challenges may be addressed, whether through new tools and techniques, new technologies, new methodologies, or new groups collaborating in the process
	
	We believe that SE research has tremendous potential to benefit scientific computing, and we have identified three key ways in which the CSE community could leverage that potential: (1) by translating existing SE research techniques and findings to the CSE domain, (2) by partnering with SE researchers to conduct novel research in established topic areas, and (3) by taking a leading role in defining SE research in emerging domains of interest to the DOE.
	
	First, SE research provides a rich toolkit of approaches for studying both software (\textit{e.g.,} program analysis and repository data mining) and those who create and use that software (\textit{e.g.,} case studies and controlled experiments). These techniques have been successfully applied to many of the notional topics in the ASCR workshop call including the impact of static and dynamic tools on software development, models of developer communities, and strategies for addressing software reliability. A promising course of action would be to build upon existing SE literature to address the same concerns in scientific software development. As we mentioned previously, the relative scarcity of published studies targeting scientific software teams could limit the transferability of findings. We see this as an opportunity to conduct translational research, such as replication studies, to fill that gap. Doing so would promote a deeper understanding of the best available evidence on the science of software development, spur closer collaboration with academia, and raise the visibility of the CSE community in SE research.
	
	Second, there are areas in SE research which are well-established but where we know scientific software developers face issues that do not neatly align with those in conventional software development. As noted by Carver et al., scientific software teams frequently encounter issues in areas such as verification and validation, performance engineering, and the use of common development tools (like IDEs)~\cite{carver2007software}. Likewise, we see broader needs that SE research could help satisfy, such as in developing strategies for managing scientific software evolution, development methodologies tailored to scientific software development, models for scientific software security, and data-driven approaches to scientific software ecosystem sustainability. Through partnership with SE researchers	-- giving them the opportunity to apply their knowledge to a new domain -- we could real progress in these fronts.
	
	Finally, looking towards the horizon of scientific computing, we anticipate increasing heterogeneity and diversity in computing architectures (\textit{e.g.,} neuromorphic, optical, quantum, etc.) as well as novel applications and use cases for HPC scientific computing (\textit{e.g.,} machine learning, hybrid edge computing). The SE community has interests in all of these subjects, but research is still nascent, such as in SE for machine learning\cite{khomh2018software} and quantum computing\cite{piattini2021quantum}. The national labs are well-positioned to take a leadership role in defining evidence-based best practices through SE research in these emerging areas.

	\section{Why Now?}

	The growing diversity of computational science teams has drawn software engineering researchers and practitioners into the fold. This is of immediate practical benefit as these professionals are bringing with them much-needed insights from other software domains to scientific computing. Case in point, at Sandia National Laboratories we have made some strides towards bringing software science into the conversation with the establishment of the Department of Software Engineering and Research within the Center for Computing Research in late 2018~\cite{milewicz2020research,willenbring2020moving}. As we turn our attention to the future, we see an opportunity to re-imagine the relationship between computational science and software engineering research; by expanding the workforce to include more software scientists, we gain new, native connections to the field. It is our sincere hope that this marks the beginning of an enduring partnership, one in which software engineering empowers computational science.

\bibliographystyle{IEEEtran}
\bibliography{ascr_softwarescience.bib}

\end{document}